\documentclass{elsarticle}

\usepackage{lineno,hyperref}
\usepackage{amsmath}
\usepackage{amssymb}
\usepackage{caption}
\usepackage{comment}
\usepackage{color}
\usepackage{lscape}
\usepackage{booktabs}
\usepackage{multirow}
\usepackage{multicol}
\usepackage{mathrsfs}
\usepackage{bm}
\usepackage{subfigure}

\journal{Icarus}

\usepackage{numcompress}
\biboptions{authoryear}

\begin{document}

\begin{frontmatter}

  \title{Mixing model of Phobos' bulk elemental composition for the determination of its origin: Multivariate analysis of MMX/MEGANE data}
  \date{November 6, 2023}

  \author[1staddress,2ndaddress]{Kaori Hirata\corref{mycorrespondingauthor}}
  \ead{hirata-kaori444@g.ecc.u-tokyo.ac.jp}
  \cortext[mycorrespondingauthor]{Corresponding author}

  \author[1staddress]{Tomohiro Usui}
  \author[1staddress]{Ryuki Hyodo}
  \author[3rdaddress]{Hidenori Genda}
  \author[1staddress]{Ryota Fukai}
  \author[4thaddress]{David J. Lawrence}
  \author[4thaddress]{Nancy L. Chabot}
  \author[4thaddress]{Patrick N. Peplowski}
  \author[5thaddress]{Hiroki Kusano}

  \address[1staddress]{Institute of Space and Astronautical Science (ISAS), Japan Aerospace Exploration Agency (JAXA), 3-1-1 Yoshinodai, Sagamihara, Kanagawa 2525210, Japan}
  \address[2ndaddress]{Department of Earth and Planetary Science, The University of Tokyo, 7-3-1 Hongo, Bunkyo, Tokyo 1130033, Japan}
  \address[3rdaddress]{Earth-Life Science Institute (ELSI), Tokyo Institute of Technology, 2-12-1 Ookayama, Meguro, Tokyo,1528550, Japan}
  \address[4thaddress]{The Johns Hopkins University Applied Physics Laboratory, Laurel, MD 20723, USA}
  \address[5thaddress]{National Institutes for Quantum Science and Technology, 4-9-1 Anagawa, Inage, Chiba 2638555, Japan}

  \begin{abstract}
    The formation process of the two Martian moons, Phobos and Deimos, is still debated with two main competing hypotheses: the capture of an asteroid or a giant impact onto Mars. In order to reveal their origin, the Martian Moons eXploration (MMX) mission by Japan Aerospace Exploration Agency (JAXA) plans to measure Phobos' elemental composition by a gamma-ray and neutron spectrometer called MEGANE. This study provides a model of Phobos' bulk elemental composition, assuming the two formation hypotheses. Using the mixing model, we established a MEGANE data analysis flow to discriminate between the formation hypotheses by multivariate analysis. The mixing model expresses the composition of Phobos in 6 key lithophile elements that will be measured by MEGANE (Fe, Si, O, Ca, Mg, and Th) as a linear mixing of two mixing components: material from Mars and material from an asteroid as represented by primitive meteorite compositions. The inversion calculation includes consideration of MEGANE's measurement errors ($E_P$) and derives the mixing ratio for a given Phobos composition, based on which the formation hypotheses are judged.
    For at least 65\% of the modeled compositions, MEGANE measurements will determine the origin uniquely ($E_P$ = 30\%), and this increases from 74 to 87\% as $E_P$ decreases from 20 to 10\%. Although the discrimination performance depends on $E_P$, the current operation plan for MEGANE predicts an instrument performance for $E_P$ of 20–-30\%, resulting in ~70\% discrimination between the original hypotheses. MEGANE observations can also enable the determination of the asteroid type of the captured body or the impactor. The addition of other measurements, such as MEGANE's measurements of the volatile element K, as well as observations by other MMX remote sensing instruments, will also contribute to the MMX mission's goal to constrain the origin of Phobos.
  \end{abstract}

\end{frontmatter}

% \linenumbers

\section{Introduction}
\label{Introduction}
The study of the Mars-moons system is crucial for understanding the initial environment of Mars as seen in the studies of the Earth-Moon system. The Martian moons, Phobos and Deimos, have been studied by telescope observations or remote sensing by Mars exploration missions. However, the origin of the Martian moons still remains controversial with two leading hypotheses.

One leading hypothesis is the capture of an asteroid, where an asteroid formed some distance from Mars and was captured by the gravity of Mars to become a satellite. This hypothesis is mainly supported by the similarity of surface characteristics (\citet{pollack_phobos_1977}; \citet{thomas_satellites_1992}) and surface spectra (\citet{burns_dynamical_1978}; \citet{burns_contradictory_1992}; \citet{pajola_phobos_2013}; \citet{pollack_gas_1979}) between the Martian moons and main-belt asteroids. The surface spectra of Phobos and Deimos are characterized by their low albedo and spectral properties similar to D-type asteroids that lack a diagnostic absorption band (\citet{murchie_color_1991}; \citet{pang_composition_1978}; \citet{pollack_multicolor_1978}; \citet{rivkin_near-infrared_2002}). In contrast to the spectral similarity, the observed orbital properties of Phobos and Deimos are difficult to account for by the capture origin. The capture origin predicts a high eccentricity and high inclination of the moons' initial orbits, which is inconsistent with the present orbits of the Martian moons (\citet{safronov_accumulation_1977}).
Numerical models have examined the processes to change their orbits after the gravitational capture by Mars (\citet{burns_contradictory_1992}; \citet{cazenave_evolution_1980}; \citet{craddock_origin_1994}; \citet{craddock_are_2011}; \citet{hunten_capture_1979}; \citet{lambeck_orbital_1979}; \citet{szeto_orbital_1983}; \citet{rosenblatt_origin_2011}), but neither of them successfully reconstructed their orbits completely.

The second hypothesis is the in-situ formation from a circum-Martian disk produced by a giant impact. This scenario is consistent with the near-circular and near-equatorial orbits of Phobos and Deimos (\citet{canup_origin_2018}; \citet{citron_formation_2015}; \citet{craddock_are_2011}; \citet{hesselbrock_ongoing_2017}; \citet{hyodo_impact_2017-1}; \citet{hyodo_impact_2017}). The disk materials and the resultant Martian moons are expected to consist of both the impactor material and the ejecta launched from Mars by the giant impact (\citet{pignatale_impact_2018}; \citet{hyodo_impact_2017}). % \citet{rosenblatt_accretion_2016}). 
The thermophysical property of disk materials depends on the impact condition, such as the impactor size and velocity. Moreover, the giant impact hypothesis predicts the depletion of volatile elements due to the impact heating (\citet{craddock_are_2011}; \citet{hyodo_impact_2017}; \citet{nakajima_inefficient_2018}; \citet{pignatale_impact_2018}).

Other hypotheses have been proposed for the Martian moons as well, such as in-situ formation from a debris disk around Mars with the moons forming as second-generation objects (\citet{patzold_phobos_2014}). \citet{bagheri_dynamical_2021} proposed a hypothesis that a single Martian moon was tidally disrupted and split into two moons although \citet{hyodo_challenges_2022} later theoretically investigated the orbital evolution and argued that two moons that originated by splitting from a common parent body were likely to be disrupted by collisions, which is inconsistent with the existence of the current Martian moons. The compositions expected for these scenarios are similar to the main two hypotheses, depending on if the material in the debris disk or disrupted body derived fundamentally from a captured object or from Martian material. Thus, determining between these two compositional endmembers is key for determining the origin of the Martian moons.

Bulk elemental compositions reflect the moons' formation processes and potentially discriminate them.
The bulk composition is estimated as chondritic for the capture scenario, whereas it represents a mixture of chondritic and Martian materials for the impact scenario (\citet{hyodo_impact_2017}; \citet{pignatale_impact_2018}).
While the surface composition of the Martian moons likely experienced some post-formation modifications due to processes, such as late accretion and space weathering, the bulk composition could have survived those processes and preserved the original information of the building blocks (\citet{hyodo_transport_2019}; \citet{ramsley_mars_2013}; \citet{ramsley_origin_2013}; \citet{ramsley_stickney_2017}).

Japan Aerospace Exploration Agency (JAXA) is planning the Martian moons' sample return mission (MMX: Martian Moons eXploration) (\citet{kawakatsu_preliminary_2023}; \citet{kuramoto_martian_2022}; \citet{usui_importance_2020}; \citet{nakamura_science_2021}). MMX has two major science goals: 1) to reveal the origin of Martian moons and gain a better understanding of the planetary formation and material transport in the solar system, and 2) to observe processes that have an impact on the evolution of the Mars system.
To achieve these goals, MMX will conduct comprehensive mineralogical (visible to near IR imaging), geochemical (elemental abundances), and geophysical (shape and gravity) measurements by seven science payloads and analyses of returned samples of Phobos (\citet{kuramoto_martian_2022}; \citet{nakamura_science_2021}).

Among the MMX science payload is the Mars-moon Exploration with GAmma rays and NEutrons (MEGANE) instrument (\citet{lawrence_measuring_2019}).
MEGANE will use gamma-ray and neutron spectroscopy to measure the elemental composition of Phobos from orbit. By detecting gamma-rays with specific energies and neutron fluxes, MEGANE will measure the abundance of major and minor elements (e.g., O, Si, Mg, Ca, and Fe), radioactive elements (e.g., K, Th, and U), and light elements (e.g., H) on the top $ \sim30 \mathrm{cm}$ of the surface of Phobos; Measurements by OROCHI (Optical RadiOmeter composed of CHromatic Imagers; \citet{kameda_design_2021}) and MIRS (MMX InfraRed Spectrometer; \citet{barucci_mirs_2021}), which are other MMX payload instruments, will provide mineralogical and geophysical information by investigating the topmost surface ($ \sim 1 \mathrm{\mu m}$) of Phobos. Thus, MEGANE observations are expected to reveal the composition of Phobos' near-surface materials and be complemented by observations by other MMX instruments.

The elemental composition acquired by MEGANE will provide insights into the formation scenario of the Martian moons. The large spatial footprints of MEGANE will be combined to determine Phobos' average surface composition, revealing the bulk elemental composition of Phobos. Note that MEGANE's observation error depends on the observation conditions, such as the accumulation period and the orbital altitude during the observations (\citet{chabot_megane_2021}; \citet{lawrence_measuring_2019}). \citet{peplowski_global_2016} previously suggested that observations at one target-body radius for more than 10 days are needed to obtain adequate signal-to-background.
Since the composition of Phobos reflects both a formation process and the evolutionary conditions experienced by the materials (e.g., the composition of building blocks and/or thermophysical properties in the impact-induced disk), the accurate interpretation of MEGANE data to confine the formation scenario (i.e., capture versus impact) requires a comprehensive investigation under a wide range of parameters that consider the endmember compositions as well as their mixing ratios.

This study aims to establish an elemental composition model of the Martian moons applicable to interpreting the MEGANE data to discriminate among the proposed origins of the Martian moons. We constructed a mixing model of the elemental composition of the Martian moons assuming the mixing of end-components of chondritic and Martian compositions. Consideration of several types of errors were included and revealed the relationship between the plausible formation scenarios and the ability of MEGANE data to discriminate among the hypotheses. Using this model, we investigated MEGANE's discrimination performance as applied to the two main formation hypotheses for the Martian moons.

\section{Method}
\label{Method}
\begin{figure}
  \centering
  \includegraphics[bb=0 0 220 222]{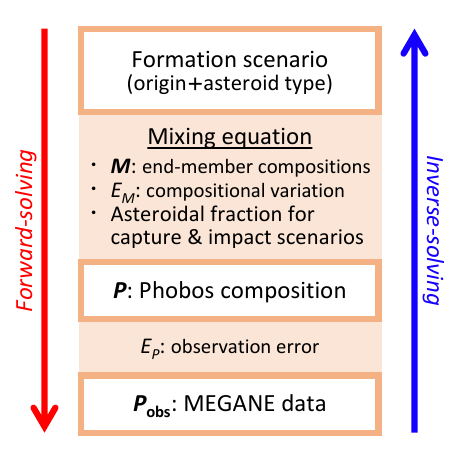}
  \caption{The forward- and inverse-solving approaches using the mixing model. This model connects the formation scenarios of Phobos and the elemental composition that will be measured by MMX MEGANE.}
  \label{fig:method-summary}
\end{figure}

This study constructed a model for Phobos' elemental composition that connects the formation scenarios proposed and the elemental composition that will be measured by MEGANE assuming a mixture of the two end-components of Martian and asteroidal materials (Fig.~\ref{fig:method-summary}). First, the forward-solving approach, which predicts the composition from each origin scenario, is introduced (Section \ref{Mixing Model: Forward-solving Approach}). Second, the inverse-solving approach to discriminate among the origin scenarios from MEGANE measurements is shown (Section \ref{Discrimination of the Origin: Inverse-solving Approach}). Finally, we define the discrimination performance to evaluate MEGANE's ability to distinguish among the origin hypotheses and investigate the dependency on parameters that are related to MEGANE's operations and measurements (Section \ref{Discrimination Performance}).

\subsection{Mixing Model: Forward-solving Approach}
\label{Mixing Model: Forward-solving Approach}

\subsubsection{Concept}
\label{Concept}

\begin{figure}
  \centering
  \includegraphics[bb=0 0 313 207]{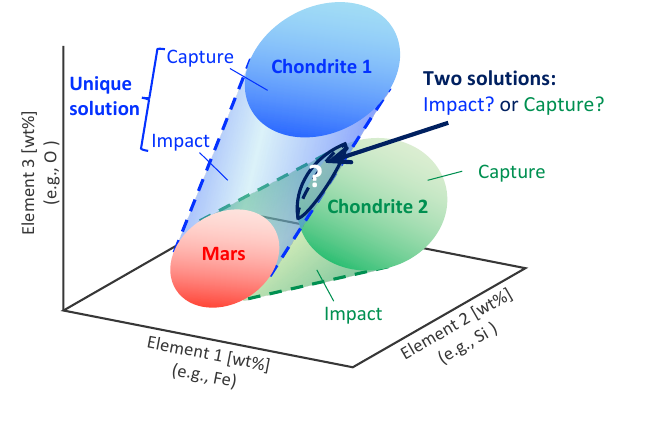}
  \caption{A schematic of the mixing model for Phobos' elemental composition. Spheroids represent end-member compositions with compositional variations (red: Martian composition, blue and green: asteroidal composition). In the case of a capture origin, Phobos' composition is modeled as that of the captured asteroid composition (i.e., within the blue or green spheres). On the other hand, in the case of an impact origin, Phobos' composition should be an intermediate composition between Mars and an asteroid (i.e., between the two red and blue or green spheres).}
  \label{fig:mixing-model}
\end{figure}

We defined the mixing model for the composition of Phobos based on the two main formation hypotheses (Fig.~\ref{fig:mixing-model}). The composition of Phobos was expressed as representing a two-component mixture with a certain mixing ratio ($r\%$) between the Martian composition ($100-r\%$) and an asteroid composition ($r\%$). Note that $r = 100\%$ in the case of the capture origin and can be a range of values between $0\%$ and $100\%$ in the case of the impact origin.
This model \textit{forwardly} predicted Phobos' composition and measurements that would be obtained from MEGANE's observation data for a given formation scenario (formation hypothesis + asteroid type).

\subsubsection{Parameters and Assumptions}
\label{Parameters and Assumptions}
To illustrate a variety of Phobos' origin scenarios, our model used 3 parameters: the composition of mixing end-members, the modeled asteroidal fraction for capture and impact origins, and the MEGANE's observation error.

\paragraph{The composition of mixing end-members}
\label{The composition of mixing end-members}
Meteorite data were used for the mixing end-member compositions using Martian and asteroidal compositions (Table \ref{tab:end-comp}).
For the Mars component, a composition for a silicate portion (Bulk Silicate Mars, BSM; \citet{taylor_bulk_2013}) was assumed in our model. %BSMを採用した->地殻ではない理由->地殻を考慮しても結果は変わらない
This is based on the calculation by \citet{hyodo_impact_2017} which indicated that the Martian ejecta in the impact origin scenarios would mainly come from a depth where both the crust and mantle were included and the compositions were not dominated by the crustal portion alone. Nevertheless, we would like to note that consideration of the diversity of crustal composition on Mars (e.g., \citet{hahn_mars_2007}) did not change our results because variations among Martian compositions are relatively small compared with the compositional differences among Mars and asteroidal compositions.
For the asteroid end-members, chondritic compositions that are considered to correspond to major components of main-belt asteroids (\citet{demeo_solar_2014}; \citet{gradie_compositional_1982}), i.e., S-, C-, E-, and D-type asteroids, were applied (Table \ref{tab:end-comp}). Chondrites are traditionally classified into 3 groups, Carbonaceous Chondrite (CC), Ordinary Chondrite (OC), and Enstatite Chondrite (EC) (\citet{brearley_chondritic_1998}). Each class is further composed of several groups. %We selected the meteorite groups within CC, OC, and EC class because of their primitive compositions and Tagish Lake because of the similarity to D-type composition. 
In this study, elemental abundances of eleven chondrite groups with primitive compositions and one ungrouped chondrite with a composition similar to D-type asteroids (\citet{alexander_quantitative_2019}; \citet{alexander_quantitative_2019-1}) were used: 6 CCs (CI, CM, CO, CV, CK, and CR), 3 OCs (H, L, and LL), 2 ECs (EH and EL), and 1 ungrouped (Tagish Lake) (Table \ref{tab:end-comp}). %This study does not consider non-chondritic compositions, which are related to E-, V-, or M-type asteroid, as end-member components because they are less likely to be captured by or collide with Mars due to thier smaller number fraction around the Mars orbit (\citet{gradie_compositional_1982}).

MEGANE can measure the abundance of major elements (e.g., Fe, Si, O, Ca, and Mg) and radioactive elements (e.g., K and Th) (\citet{lawrence_measuring_2019}). This study followed the element classification adopted by \citet{taylor_bulk_2013}, in which Fe, Si, O, Ca, Mg, and Th are referred to as lithophile elements and K is classified as a moderately volatile element. Among these elements, we selected 6 lithophile elements to model Phobos' composition. In Section \ref{The effect of volatile loss}, we discuss the use of K abundance as well.

Additionally, this study took into account the end-members' compositional variations and introduced a relative error of 10\%.

\begin{landscape}
  \begin{table}[h]
    \centering
    \begin{tabular}{cccccccccccccc}
      \toprule
                & BSM$^a$ & CI$^b$ & CM$^b$ & CO$^b$ & CV$^b$ & CK$^b$ & CR$^b$ & H$^c$ & L$^c$ & LL$^c$ & EH$^c$ & EL$^c$ & TL$^b$ \\
      \midrule
      Fe [wt\%] & 14.1    & 18.5   & 21.2   & 24.7   & 23.6   & 23.4   & 24.1   & 27.5  & 21.5  & 18.5   & 29.0   & 22.0   & 18.9   \\
      Si [wt\%] & 20.5    & 10.7   & 13.0   & 16.0   & 16.0   & 15.9   & 15.8   & 16.9  & 18.5  & 18.9   & 16.7   & 18.6   & 11.6   \\
      O [wt\%]  & 42.0    & 45.9   & 40.4   & 35.1   & 36.7   & 37.0   & 36.8   & 32.9  & 36.3  & 37.7   & 28.1   & 33.0   & 37.9   \\
      Ca [wt\%] & 1.74    & 0.91   & 1.24   & 1.56   & 1.85   & 2.00   & 1.38   & 1.25  & 1.31  & 1.30   & 0.85   & 1.08   & 0.99   \\
      Mg [wt\%] & 18.5    & 9.5    & 11.7   & 14.4   & 14.8   & 14.8   & 13.9   & 14.0  & 14.9  & 15.3   & 10.6   & 14.1   & 10.9   \\
      Th [ppb]  & 5.8     & 3.0    & 3.9    & 4.5    & 6.3    & 5.8    & 4.2    & 4.2   & 4.3   & 4.3    & 3.0    & 3.4    & 3.8    \\
      \bottomrule
    \end{tabular}
    \caption{The elemental composition of 6 elements (Fe, Si, O, Ca, Mg, and Th) for the 13 end-members (Mars (BSM; Bulk Silicate Mars), 6 CCs, 3 OCs, 2 ECs, and ungrouped (TL; Tagish Lake)) assumed in the model.\\
      $^a$: \citet{taylor_bulk_2013}, $^b$: \citet{alexander_quantitative_2019-1}, $^c$: \citet{alexander_quantitative_2019}}
    \label{tab:end-comp}
  \end{table}
\end{landscape}

\paragraph{Modeled asteroidal fraction for capture and impact origins}
\label{Modeled asteroidal fraction for capture and impact origins}
In the case of the asteroid capture hypothesis, the elemental composition of Phobos is similar to that of a captured asteroid. In this case, our model assumed that Phobos' building blocks are composed only of the material from the captured asteroid, resulting in an asteroidal fraction of 100\%.

On the other hand, in the giant impact hypothesis, the fraction of impactor material (i.e., modeled asteroidal fraction) varies depending on the impact condition, such as the size of the impactor or the impact angle and velocity (e.g.,\citet{canup_origin_2018}; \citet{hyodo_impact_2017}). Using numerical simulations, \citet{hyodo_impact_2017} investigated the thermophysical properties of the impact-induced disk from which the Martian moons formed. They suggested that the building blocks of Phobos should contain both Martian materials of $\gtrsim 50\%$ and impactor materials of $\gtrsim 35\%$, while the mixing ratio changes depending on the impact conditions, e.g., impact velocity or angle.
For example, disk materials are composed of 40\% Martian materials and 60\% impactor materials to form the Borealis basin on Mars with an impactor mass of 0.03 times that of Mars and an impact angle of 45°. Considering that Phobos and Deimos accreted in the outer part of the impact-induced disk (\citet{rosenblatt_accretion_2016}), 70\% of the outer disk material was estimated to come from Mars. It was also suggested that the mixing ratio of impactor material would also depend on the impact angle, changing the mixing ratio from 30\% to 65\%.
As a reference, this study assumed the modeled asteroidal fraction of 50\% for impact origin (reference case), which means that a giant impact results in a mixing of 50\% asteroidal materials and 50\% Martian materials. More practically, the uncertainty of mixing conditions was taken into account and the modeled asteroidal fraction of 30--70\% was adopted for the impact origin (practical case). %In this case, the mixing ratio should agree with the range.

\paragraph{MEGANE observation error}
\label{MEGANE observation error}
The observation error of MEGANE depends on the observation sequence, especially the accumulation period and the orbital altitude. \citet{peplowski_global_2016} previously suggested that gamma-ray and neutron measurements require orbital altitudes less than or equal to 1-target body radius for successful analysis. \citet{lawrence_measuring_2019} estimated the observation error needed to meet the MEGANE science objectives element by element and determined that this error could be achieved with at least 10 days of accumulation at altitudes equal to or less than 1-target body radius. We assumed 30, 20, 10, and 0\% relative error for the MEGANE observation error in our model calculations: $E_P = 30$, $20$, $10$, and $0$ [\%].

\subsubsection{Mixing Equation}
Our mixing model calculated the composition of Phobos as a linear sum of Martian and asteroidal compositions, with a certain mixing ratio.
Phobos' composition ($\bm{P}$) resulting from the mixing of the compositions of Mars ($M_0$) and an asteroid $i$ ($M_i$; $i= 1,2, \cdots, 12$) was expressed by Eq.~\ref{eq:mixeq_mat}, using matrices $\bm{P}$, $\bm{M}$, and $\bm{R}$. $\bm{P}$ represented the abundance of 6 elements on Phobos: Fe, O, Si, Ca, Mg, and Th. The end-member composition matrix $\bm{M}$ was composed of the same 6-element composition of Mars ($M_0$) and asteroids ($M_i$; $i= 1,2, \cdots, 12$), $\bm{M}=[M_0 M_1 \cdots M_{12}]$. The mixing ratio matrix $\bm{R}$ was composed of the mixing ratios for $i$-th end-member compositions ($r_i$; $i= 0,1,2, \cdots, 12$).
Note that since we assumed the mixing of only two end-components, i.e., Mars and the selected type of asteroid $i'$, $r_i=0$ for $i \ne i'$ and $r_0 = 1-r_{i'}$. The abundance of each element ($P_e$; $e=$ Fe, O, Si, Ca, Mg, and Th) in Phobos' material was written down using that in Mars ($M_{e,0}$) and asteroid ($M_{e,i}$) materials and the mixing ratio $r_i$. The subscript $e$ indicates the type of elements (Fe, O, Si, Ca, Mg, and Th), and $i$ indicates the mixing end-members (Martian component for $i = 0$ and asteroids for $i=1$--$12$).
\begin{eqnarray}
  %\begin{equation}
  \bm{P}&=&\bm{M}\bm{R}, \nonumber\\
  %\end{equation}
  %\begin{equation}
  \label{eq:mixeq_mat}
  \begin{pmatrix}
    P_{\mathrm{Fe}} \\
    P_{\mathrm{O}}  \\
    P_{\mathrm{Si}} \\
    P_{\mathrm{Ca}} \\
    P_{\mathrm{Mg}} \\
    P_{\mathrm{Th}} \\
  \end{pmatrix}
  &=&
  \begin{pmatrix}
    M_{\mathrm{Fe},0} & M_{\mathrm{Fe},1} & \cdots & M_{\mathrm{Fe},12} \\
    M_{\mathrm{O},0}  & M_{\mathrm{O},1}  & \cdots & M_{\mathrm{O},12}  \\
    M_{\mathrm{Si},0} & M_{\mathrm{Si},1} & \cdots & M_{\mathrm{Si},12} \\
    M_{\mathrm{Ca},0} & M_{\mathrm{Ca},1} & \cdots & M_{\mathrm{Ca},12} \\
    M_{\mathrm{Mg},0} & M_{\mathrm{Mg},1} & \cdots & M_{\mathrm{Mg},12} \\
    M_{\mathrm{Th},0} & M_{\mathrm{Th},1} & \cdots & M_{\mathrm{Th},12} \\
  \end{pmatrix}
  \left(
  \begin{matrix}
      r_0    \\
      r_1    \\
      \vdots \\
      r_i    \\
      \vdots \\
      r_{11} \\
      r_{12} \\
    \end{matrix}
  \right).
  %\end{equation}
\end{eqnarray}
\begin{comment}
\begin{matrix}
  1-r_i  \\
  0      \\
  \vdots \\
  0      \\
  r_i    \\
  0      \\
  \vdots \\
  0      \\
\end{matrix}
\end{comment}
Considering relative errors $E_P$ for $\bm{P}$ and $\bm{M}$ for $E_M$, $\bm{P_{\mathrm{obs}}}$ and $\bm{M}$ should be included in the range of $[\bm{P_\mathrm{obs,min}}:\bm{P_\mathrm{obs,max}}]$ and $[\bm{M_\mathrm{min}}:\bm{M_\mathrm{max}}]$, respectively, which were given as
\begin{eqnarray}
  \bm{P_\mathrm{obs,min}}=(P_{\mathrm{obs,min},e})=\bm{P} \times \frac{100-E_P}{100}, \\
  \bm{P_\mathrm{obs,max}}=(P_{\mathrm{obs,max},e})=\bm{P} \times \frac{100+E_P}{100}, \\
  \bm{M_\mathrm{min}}=(M_{\mathrm{min},e,i})=\bm{M} \times \frac{100-E_M}{100}, \\
  \bm{M_\mathrm{max}}=(M_{\mathrm{max},e,i})=\bm{M} \times \frac{100+E_M}{100}.
\end{eqnarray}

\subsection{Discrimination of the Origin: Inverse-solving Approach}
\label{Discrimination of the Origin: Inverse-solving Approach}
\begin{figure}
  \centering
  \includegraphics[bb=0 0 370 280]{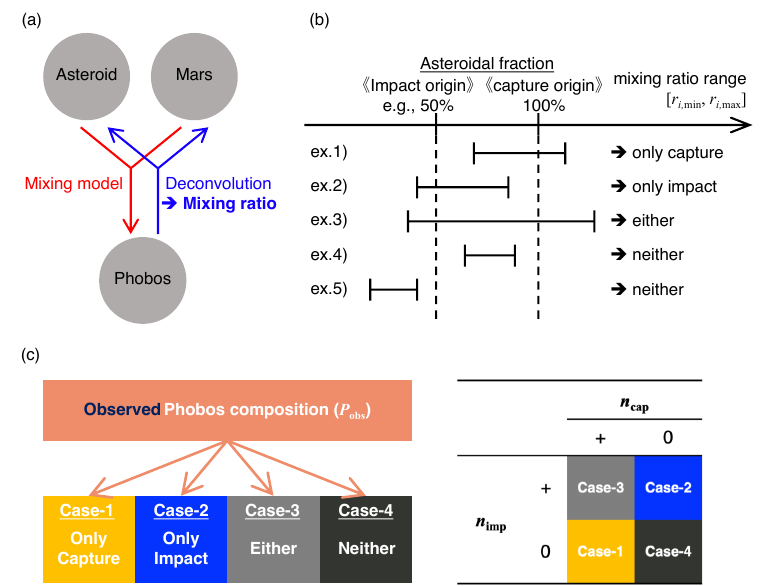}
  \caption{Schematics showing 3 calculation steps in the inverse-solving approach. (a) The observed Phobos composition is deconvolved into two mixing end-members. The mixing ratio is calculated (Section \ref{Mixing ratio calculation}). (b) Formation hypotheses are judged based on whether the derived mixing ratio range agrees with the modeled asteroidal fraction for capture and impact origins. These calculations are performed on all types of asteroid compositions independently (Section \ref{Criteria for capture/impact hypothesis}). (c) By summarizing the judgments for all the asteroid types, a given Phobos composition measured by MEGANE is classified into 4 cases (Section \ref{Classification based on the reasonable formation scenario}).}
  \label{fig:Calculation-flow}
\end{figure}

The \textit{inverse}-solving approach uses our model to determine the origin of Phobos from its composition by using MEGANE data. First, a given composition was deconvolved into two mixing end components. The inverse calculation derived the mixing ratio $r$ (Section \ref{Mixing ratio calculation}; Fig.~\ref{fig:Calculation-flow}(a)).
Next, the derived mixing ratio judged whether the formation scenarios were possible to explain the composition or not, based on the criteria (Section \ref{Criteria for capture/impact hypothesis}; Fig.~\ref{fig:Calculation-flow}(b)).
By summarizing the judgments for all asteroid types, the composition was classified into 4 cases (Section \ref{Classification based on the reasonable formation scenario}; Fig.~\ref{fig:Calculation-flow}(c)).

\subsubsection{Mixing ratio calculation}
\label{Mixing ratio calculation}
The mixing ratio for a given MEGANE compositional determination and a given asteroid type was derived from inverse calculations of the mixing equation (Eq.~\ref{eq:mixeq_mat}). The mixing ratio for each element $r_{e,i}$ was determined.
\begin{comment}
\begin{equation}
  \label{eq:P_e}
  P_e=M_{e,0} (1-r_{e,i})+M_{e,i} r_{e,i}
\end{equation}
\begin{equation}
  \label{eq:r_e,i}
  r_{e,i}=\frac{M_{e,0}-P_e}{M_{e,0}-M_{e,i}}
\end{equation}
\end{comment}
Since we assumed the MEGANE observation error $E_P$, the error for $\bm{P}$, and the compositional variation of Mars and asteroids $E_M$, the error for $\bm{M}$, the minimum and maximum values of $r_{e,i}$ ($r_{e,i,min}$ and $r_{e,i,max}$) were calculated as %(Eqs.~\ref{eq:r_e,i,min}--\ref{eq:r_e,i,max}). 
\begin{eqnarray}
  \label{eq:r_e,i,min}
  r_{e,i,\mathrm{min}}=
  \begin{cases}
    \frac{M_{\mathrm{min},e,0}-P_{\mathrm{obs,max},e}}{M_{\mathrm{max},e,0}-M_{\mathrm{max},e,i}} & (M_{e,0} \leqq M_{e,i}), \\
    \frac{M_{\mathrm{min},e,0}-P_{\mathrm{obs,min},e}}{M_{\mathrm{min},e,0}-M_{\mathrm{min},e,i}} & (M_{e,0} \geqq M_{e,i}), \\
  \end{cases}  \\
  \label{eq:r_e,i,max}
  r_{e,i,\mathrm{max}}=
  \begin{cases}
    \frac{M_{\mathrm{max},e,0}-P_{\mathrm{obs,min},e}}{M_{\mathrm{min},e,0}-M_{\mathrm{min},e,i}} & (M_{e,0} \leqq M_{e,i}), \\
    \frac{M_{\mathrm{max},e,0}-P_{\mathrm{obs,max},e}}{M_{\mathrm{max},e,0}-M_{\mathrm{max},e,i}} & (M_{e,0} \geqq M_{e,i}).
  \end{cases}
\end{eqnarray}
\begin{comment}

\end{comment}
Note that $\bm{P}$ represented the Phobos composition measured by MEGANE $\bm{P_{\mathrm{obs}}}$ in the inverse approach.

From the derived set of the mixing ratio range $[r_{e,i,\mathrm{min}}, r_{e,i,\mathrm{max}}]$ for element $e$ within 6 elements, we considered the common range among the 6 elements $[r_{i,\mathrm{min}}, r_{i,\mathrm{max}}]$ as a possible solution for a given set of MEGANE observation data $\bm{P_{\mathrm{obs}}}$ and a given asteroid type $i$, as
\begin{eqnarray} % r_{i,min},r_{i,max}
  % \label{eq:r_i,min}
  % r_{i,min}=\max\limits_{e=1,6} [r_{e,i}-\Delta r_{e,i}],\\
  % \label{eq:r_i,max}
  % r_{i,max}=\min\limits_{e=1,6} [r_{e,i}+\Delta r_{e,i}].
  \label{eq:r_i_range}
  [r_{i,min},r_{i,max}]=
  % \begin{cases}
  [\max\limits_{e=1,6}[r_{e,i,\mathrm{min}}], \min\limits_{e=1,6}[r_{e,i,\mathrm{max}}]],
  % \end{cases}
\end{eqnarray}
when $\max\limits_{e=1,6}[r_{e,i,\mathrm{min}}] \le \min\limits_{e=1,6}[r_{e,i,\mathrm{max}}]$. Otherwise $[r_{i,\mathrm{min}},r_{i,\mathrm{max}}]$ would not have a solution. %have an empty range. %\\
% [0,0] & (\max\limits_{e=1,6}[r_{e,i,\mathrm{min}}] \gt \min\limits_{e=1,6}[r_{e,i,\mathrm{max}}]).To show an example, when $[r_{\mathrm{Fe},i,\mathrm{min}}, r_{\mathrm{Fe},i,\mathrm{max}}]=[40\%,60\%]$ and $[r_{\mathrm{O},i,\mathrm{min}}, r_{\mathrm{O},i,\mathrm{max}}]=[50\%,70\%]$, $[r_{i,\mathrm{min}}, r_{i,\mathrm{max}}]=[50\%,60\%]$.

\subsubsection{Criteria for capture/impact hypothesis}
\label{Criteria for capture/impact hypothesis}
The derived mixing ratio range $[r_{i,\mathrm{min}}, r_{i,\mathrm{max}}]$ was used to judge the origin based on the modeled asteroidal fractions for the two formation hypotheses (Section \ref{Parameters and Assumptions}; Fig.~\ref{fig:Calculation-flow}(b)). When the modeled asteroidal fraction is sandwiched between the derived mixing ratio range, the $\bm{P_{\mathrm{obs}}}$ was explained by the scenario: $\bm{P_{\mathrm{obs}}}$ could be explained by the capture and giant impact of asteroid $i$ if the modeled asteroidal fraction for capture origin (100\%) and impact origin (e.g., 50\% in reference case) is sandwiched between $[r_{i,\mathrm{min}} r_{i,\mathrm{max}}]$, respectively.

\subsubsection{Classification based on the reasonable formation scenario}
\label{Classification based on the reasonable formation scenario}
Based on the origin judgments, we counted the number of asteroid types ($n_{\mathrm{cap}}$ and $n_{\mathrm{imp}}$) that accounted for the $\bm{P_{\mathrm{obs}}}$ in the capture and impact origins, respectively (Fig.~\ref{fig:Calculation-flow}(c)). Then $\bm{P_{\mathrm{obs}}}$ was classified into 4 cases: (Case-1) in case $n_{\mathrm{cap}} > 0$ and $n_{\mathrm{imp}} = 0$, only the capture hypothesis can explain a given $\bm{P}$, (Case-2) in case $n_{\mathrm{cap}} = 0$ and $n_{\mathrm{imp}} > 0$, only the impact hypothesis can explain a given $\bm{P}$, (Case-3) in case $n_{\mathrm{cap}} > 0$ and $n_{\mathrm{imp}} > 0$, either of the two hypotheses can explain a given $\bm{P}$, and (Case-4) in case $n_{\mathrm{cap}} = 0$ and $n_{\mathrm{imp}}= 0$, neither hypothesis can explain a given $\bm{P}$. Under this definition, we can say that the origins of Phobos are determined in Case-1 or -2.

\subsection{Discrimination Performance}
\label{Discrimination Performance}
To evaluate the feasibility of the discrimination of Phobos' origin using MEGANE data and our model, we changed $\bm{P_{\mathrm{obs}}}$ within the 6-dimensional space $[\bm{P}_{\mathrm{min}}:\bm{P}_{\mathrm{max}}]$ and investigated the origin for any $\bm{P_{\mathrm{obs}}} \in [\bm{P_{\mathrm{min}}}:\bm{P_{\mathrm{max}}}]$. Modeled composition ranges for each element $P_{\mathrm{min}, e}$ and $P_{\mathrm{max}, e}$ were determined by
\begin{eqnarray}
  \label{eq:P_min}
  P_{\mathrm{min},e}=\min_{i=0,12} [M_{\mathrm{min},e,i}]
\end{eqnarray}
and
\begin{eqnarray}
  \label{eq:P_max}
  P_{\mathrm{max},e}=\max_{i=0,12} [M_{\mathrm{max},e,i}].
\end{eqnarray}

We defined ``modeled compositions'' as all $\bm{P}$ that can be explained by the capture and/or impact hypotheses: Case-1, -2, or -3. ``Hypothesis-discriminating compositions'' were also defined as $\bm{P}$ that can be explained only by a unique formation hypothesis: Case-1 or -2.

To evaluate our 6-dimensional results, discrimination performance was defined as the ratio of hypothesis-discriminating compositions and modeled compositions. %(Eqs.~\ref{eq:D}--\ref{eq:D_imp}). 
$\mathscr{D}$, $\mathscr{D}_{\mathrm{cap}}$, and $\mathscr{D}_{\mathrm{imp}}$ were given as
\begin{eqnarray}
  \label{eq:D}
  \mathscr{D}=\mathscr{D}_{\mathrm{cap}}+\mathscr{D}_{\mathrm{imp}}=\frac{n_1+n_2}{n_1+n_2+n_3} \times 100 \mathrm{\ [\%]}, \\
  \label{eq:D_cap}
  \mathscr{D}_{\mathrm{cap}}=\frac{n_1}{n_1+n_2+n_3}\times 100 \mathrm{\ [\%]},\\
  \label{eq:D_imp}
  \mathscr{D}_{\mathrm{imp}}=\frac{n_2}{n_1+n_2+n_3}\times 100 \mathrm{\ [\%]},
\end{eqnarray}
where $n_1, n_2,$ and $n_3$ were the number of data points of $\bm{P}$ classified into Case-1, -2, and -3, respectively. $\mathscr{D}$, a sum of $\mathscr{D}_{\mathrm{cap}}$ and $\mathscr{D}_{\mathrm{imp}}$, indicated the ratio of hypothesis discriminating compositions to modeled compositions, that is, the extent to which the formation hypothesis was discriminated within modeled compositions. $\mathscr{D}_{\mathrm{cap}}$ and $\mathscr{D}_{\mathrm{imp}}$ were the ratios of $\bm{P}$ related only to the capture and impact origins, respectively.

To visualize the relationship between the compositions and origins of Phobos, we calculated the discrimination performance by 5 cases of 2-element compositions. Here the fixed two elements were pairs of Fe-, O-, Ca-, Mg-, and Th-Si, while compositions of the rest of the four elements were changed to recalculate $n_\mathrm{cap}$ and $n_\mathrm{imp}$ in Eqs.~\ref{eq:D}--\ref{eq:D_imp}. We denoted them as $d$, $d_{\mathrm{cap}}$, and $d_{\mathrm{imp}}$ to distinguish from $\mathscr{D}$, $\mathscr{D}_{\mathrm{cap}}$, and $\mathscr{D}_{\mathrm{imp}}$.

\section{Results}
\label{Results}
\subsection{Reference Case: Modeled Asteroidal Fraction for Impact Origin of 50\%}
\label{Reference case}
For the reference case, $\mathscr{D}$ were calculated with varying MEGANE observation error $E_P$ (Asteroidal fraction of 50\% (reference case)' in Table \ref{tab:Dh_cases}). $\mathscr{D}$ was 64.7\% when $E_P = 30\%$ and it increased as $E_P$ decreased. Throughout the calculated $E_P$ range (0--30\%), $\mathscr{D}_{\mathrm{cap}}$ almost agreed with $\mathscr{D}_{\mathrm{imp}}$.

\begin{figure}
  \centering
  \includegraphics[bb=0 0 226 226]{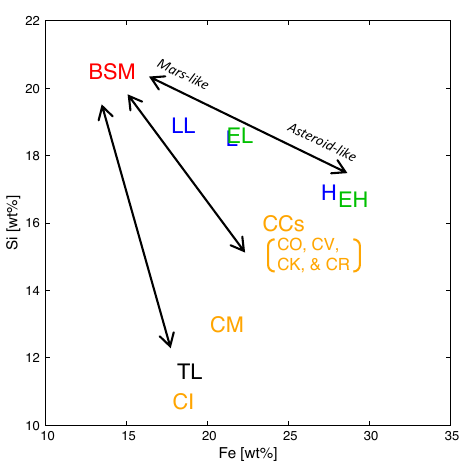}
  \caption{Variation of end-member compositions on a Fe-Si diagram (labels: Bulk Silicate Mars (BSM; red); CC (yellow); OC (blue); EC (green); and Tagish Lake (black)). End-members composition show variational extent from Mars-like composition toward asteroid-like compositions (\textit{Mars-asteroid compositional transition}; black arrows). }
  \label{fig:Two-elm_example}
\end{figure}

Here we briefly review compositional variations among end-member components (Fig.~\ref{fig:Two-elm_example}). As end-member compositions, Mars and asteroidal compositions are assigned as input parameters (Section \ref{Parameters and Assumptions}). Since we assigned several different chondritic compositions as an input parameter, the asteroidal components (yellow, blue, green, and black labels in Fig.~\ref{fig:Two-elm_example}) show a more extended distribution than the Mars component (red label in Fig.~\ref{fig:Two-elm_example}). For convenience, a transition from Mars-like to asteroid-like compositions (black arrows in Fig.~\ref{fig:Two-elm_example}) will be referred to as \textit{Mars-asteroid compositional transition} in this paper.

\begin{figure}
  \centering
  \includegraphics[bb=0 0 231 425]{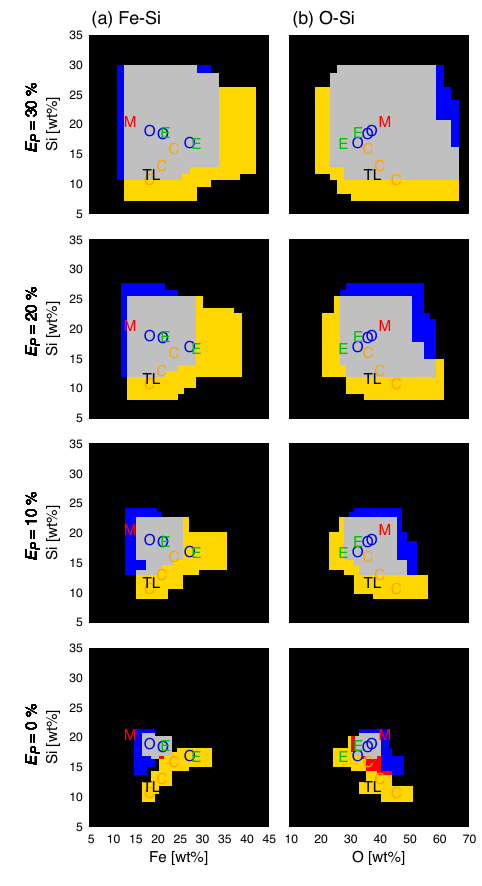}
  \caption{The relationship between 2-element compositions and the determined formation hypothesis, along with the compositional end-members (labels: Mars (``M'' in red); CC (``C'' in yellow); OC (``O'' in blue); EC (``E'' in green); and Tagish Lake (``TL'' in black)). Background colors indicate whether the formation hypothesis is identified uniquely from 2-element compositions as only capture hypothesis (yellow) or only impact hypothesis (blue), is not identified uniquely from 2-element compositions but from 6- elements compositions (red), is not identified uniquely even from 6-elements composition (gray) or is explained neither by the capture nor impact hypotheses (black). (a) Fe-, (b) O-, (c) Ca-, (d) Mg-, and (e) Th-Si diagrams for varying $E_P$ of 30, 20, 10, and 0\%.}
  \label{fig:Two-elm_colors_multi}
\end{figure}

The relationships between 6-dimensional compositions and the corresponding origins are summarized in 2-dimensional space using $d$, $d_{\mathrm{cap}}$, and $d_{\mathrm{imp}}$ (Fig.~\ref{fig:Two-elm_colors_multi}). $\bm{P}$ occurring beyond the asteroid-side of the Mars-asteroid compositional transition tended to have $d_{\mathrm{cap}}$ of 100\% (yellow in Fig.~\ref{fig:Two-elm_colors_multi}), while those occurring beyond Mars-side of the transition had $d_{\mathrm{imp}}$ of 100\% (blue in Fig.~\ref{fig:Two-elm_colors_multi}).
But if $\bm{P}$ occurred along the compositional transition, $d < 100\%$, suggesting that the formation hypothesis was not determined uniquely (gray in Fig.~\ref{fig:Two-elm_colors_multi}).

The extent of the modeled two-element compositions varied depending on sets of elements (Fig.~\ref{fig:Two-elm_colors_multi}a--e).
We compared how effectively the selected pairs of two-element compositions separate the origin, by calculating the ratios between the 2-element compositions with $d=100\%$ (yellow, blue, and red in Fig.~\ref{fig:Two-elm_colors_multi}) and modeled compositions (yellow, blue, red, and gray in Fig.~\ref{fig:Two-elm_colors_multi}).
The ratios for Fe-Si and O-Si compositions were the first and second largest among the 5 pairs, for example, 73.8 and 52.6\% when $E_P=20\%$, respectively, while Th-Si had the smallest value of 40.8\%.

Another difference is whether the compositions that $d=100\%$, $d_{\mathrm{cap}}< 100\%$, and $d_{\mathrm{imp}} < 100\%$ exist at the same time or not. These compositions determine the origins uniquely, although there are possibilities of either capture and impact origins (red in Fig.~\ref{fig:Two-elm_colors_multi}a--e). These compositions were represented only in the case of $E_P = 0\%$. Such compositions existed the most within Th-Si compositions and the least within Fe-Si compositions.

The populations of $\bm{P}$ for Case 1--4 were distributed differently for different $E_P$. Discrimination performance $\mathscr{D}$ was 64.6, 73.8, 86.6, and 95.8\% when MEGANE's error $E_P$ was 30, 20, 10, and 0\%, respectively (`Asteroidal fraction of 50\% (reference case)' in Table \ref{tab:Dh_cases}).

\begin{figure*}
  \centering
  \includegraphics[bb=0 0 335 425]{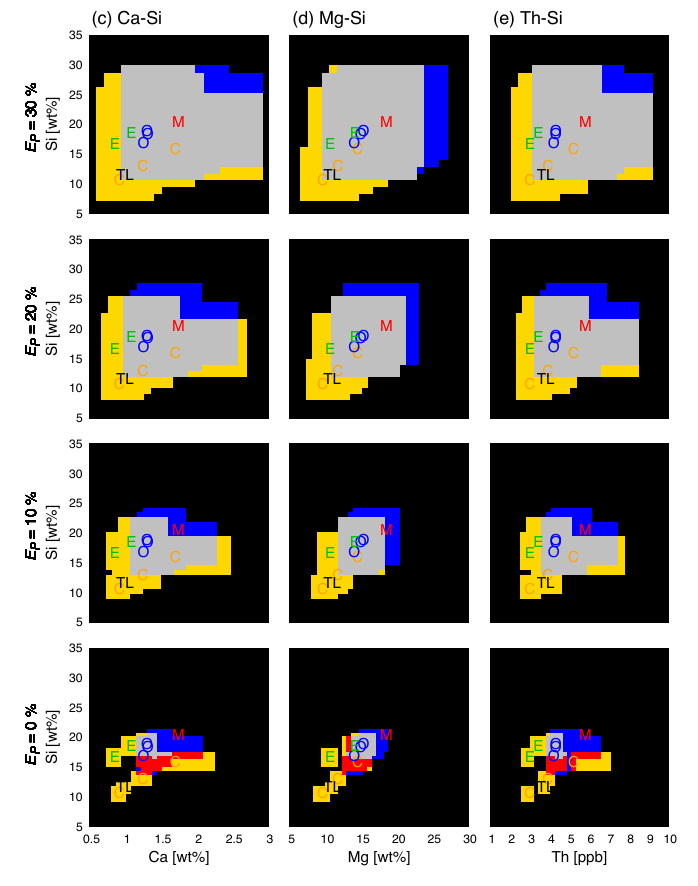}
  % \caption*{The relationship between 2-element compositions and the determined formation hypothesis. (a) An example of Fe-Si diagram for $E_P=30\%$ along with the compositional end-members (labels: Mars (red); CC (yellow); OC (blue); EC (green); and Tagish Lake (black)). The color indicates whether the formation hypothesis is identified uniquely from 2-element compositions as only capture hypothesis (yellow) or only impact hypothesis (blue), is not identified uniquely from 2-element compositions but from 6- elements compositions (red), is not identified uniquely even from 6-elements composition (gray) or is explained neither by the capture nor impact hypotheses (black). (b) Fe-, (c) O-, (d) Ca-, (e) Mg-, and (f) Th-Si diagrams for varying $E_P$ of 30, 20, 10, and 0\%.}
\end{figure*}

\begin{table}
  \centering
  \begin{tabular}{cccccccc}
    \toprule
    \multirow{3}{*}{$E_P \ [\%]$} & \multicolumn{3}{c}{\multirow{2}{*}{Asteroidal fraction of 30\%}} &                                   & \multicolumn{3}{c}{Asteroidal fraction of 50\%}                                                                                                                       \\
                                  &                                                                  &                                   &                                                     &  & \multicolumn{3}{c}{(reference case)}                                                                         \\
    \cmidrule{2-4} \cmidrule{6-8}
                                  & $\mathscr{D}$ [\%]                                               & $\mathscr{D}_{\mathrm{cap}}$ [\%] & $\mathscr{D}_{\mathrm{imp}}$ [\%]                   &  & $\mathscr{D}$ [\%]                   & $\mathscr{D}_{\mathrm{cap}}$ [\%] & $\mathscr{D}_{\mathrm{imp}}$ [\%] \\
    \midrule
    30                            & 78.3                                                             & 47.4                              & 30.9                                                &  & 64.7                                 & 34.4                              & 30.2                              \\
    20                            & 87.5                                                             & 52.2                              & 35.3                                                &  & 73.8                                 & 39.0                              & 34.8                              \\
    10                            & 95.4                                                             & 56.6                              & 38.9                                                &  & 86.6                                 & 45.3                              & 41.3                              \\
    0                             & 99.6                                                             & 52.6                              & 47.0                                                &  & 95.8                                 & 45.1                              & 50.7                              \\
    \bottomrule                                                                                                                                                                                                                                                                                                  \\

    \toprule
    \multirow{3}{*}{$E_P \ [\%]$} & \multicolumn{3}{c}{\multirow{2}{*}{Asteroidal fraction of 70\%}} &                                   & \multicolumn{3}{c}{Asteroidal fraction of 30--70\%}                                                                                                                   \\
                                  &                                                                  &                                   &                                                     &  & \multicolumn{3}{c}{(practical case)}                                                                         \\
    \cmidrule{2-4} \cmidrule{6-8}
                                  & $\mathscr{D}$ [\%]                                               & $\mathscr{D}_{\mathrm{cap}}$ [\%] & $\mathscr{D}_{\mathrm{imp}}$ [\%]                   &  & $\mathscr{D}$ [\%]                   & $\mathscr{D}_{\mathrm{cap}}$ [\%] & $\mathscr{D}_{\mathrm{imp}}$ [\%] \\
    \midrule
    30                            & 49.0                                                             & 20.8                              & 28.2                                                &  & 63.5                                 & 9.4                               & 54.1                              \\
    20                            & 60.7                                                             & 27.5                              & 33.1                                                &  & 75.5                                 & 10.0                              & 65.5                              \\
    10                            & 74.3                                                             & 35.8                              & 38.5                                                &  & 87.5                                 & 9.4                               & 78.2                              \\
    0                             & 92.4                                                             & 45.8                              & 46.6                                                &  & 97.9                                 & 4.1                               & 93.9                              \\
    \bottomrule
  \end{tabular}
  \caption{Discrimination performances ($\mathscr{D}$, $\mathscr{D}_{\mathrm{cap}}$, and $\mathscr{D}_{\mathrm{imp}}$; Eqs.~\ref{eq:C}--\ref{eq:C_imp}) calculated for modeled asteroidal fraction for impact origin of 30\%, 50\% (reference case), 70\%, and 30--70\% (practical case).}
  \label{tab:Dh_cases}
\end{table}
\begin{comment}
\begin{table}
  \centering
  \begin{tabular}{cccccccc}
    \toprule
    \multirow{2}{*}{$E_P \ [\%]$} & \multirow{2}{*}{$\mathscr{D} \ [\%]$} &  & \multicolumn{5}{c}{apparent performance $[\%]$}                                \\
    \cmidrule{4-8}
                                  &                                       &  & Fe-Si                                           & O-Si & Ca-Si & Mg-Si & Th-Si \\
    \midrule
    30                            & 64.6                                  &  & 38.0                                            & 26.6 & 30.7  & 31.8  & 30.0  \\
    20                            & 73.8                                  &  & 52.6                                            & 46.5 & 43.8  & 42.0  & 40.8  \\
    10                            & 86.6                                  &  & 63.4                                            & 54.6 & 54.4  & 53.3  & 52.7  \\
    0                             & 95.8                                  &  & 76.3                                            & 70.7 & 68.0  & 58.6  & 70.3  \\
    \bottomrule
  \end{tabular}
  \caption{Discrimination performance $\mathscr{D}$ and apparent discrimination performance for 2-elements composition (Fe-, O-, Ca-, Mg-, and Th-Si composition).}
  \label{tab:Dh_value}
\end{table}
\end{comment}

\subsection{Practical Case: Modeled Asteroidal Fraction for Impact Origin of  30--70\%}
\label{Practical case}
Since the mixing ratio may vary by 30--70\% as a function of the impact conditions, asteroidal fractions can change within that range. Here we investigated the dependency of the discrimination results on the values of asteroidal fractions.

\begin{figure}
  \centering
  \includegraphics[bb=0 0 397 556]{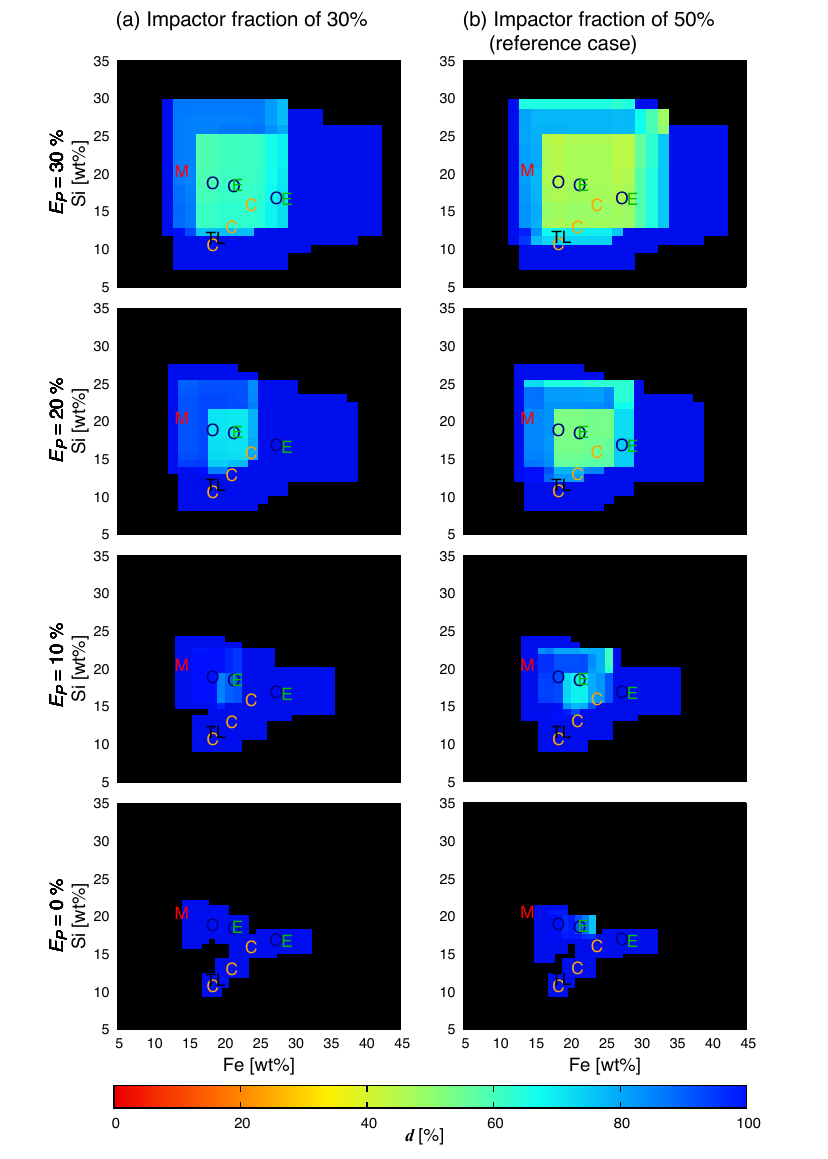}
  \caption{Discrimination performance $d$ on Fe-Si diagrams when modeled asteroidal fraction for capture origin was 100\% and for impact origin was(a) 30\%, (b) 50\% (reference case), (c) 70\%, and (d) 30-70\% (practical case), along with the compositional end-members (labels: Mars (``M'' in red); CC (``C'' in yellow); OC (``O'' in blue); EC (``E'' in green); and Tagish Lake (``TL'' in black)). Note that modeled asteroidal fractions were used as criteria when we judged the origin from the calculated mixing ratio (Section \ref{Criteria for capture/impact hypothesis}).}
  \label{fig:d_dist}
\end{figure}

When the modeled asteroidal fraction for the impact origin was 30\%, $\mathscr{D}$, $\mathscr{D}_{\mathrm{cap}}$, and $\mathscr{D}_{\mathrm{imp}}$ were improved from the reference case. In contrast, they were reduced in the case of the modeled asteroidal fraction for the impact origin of 70\% (Fig.~\ref{fig:Two-elm_colors_multi}; Table \ref{tab:Dh_cases}).
Furthermore, the transition of $\bm{P}$ for Case-2 toward Mars and asteroidal compositions were confirmed (Fig.~\ref{fig:d_dist}a--c).

As a practical case, we also assumed the modeled asteroidal fraction of 30--70\%. Under this parameter setting, it was judged as the impact origin when the derived mixing ratio agreed with any $r$ between 30--70\%. While $\mathscr{D}$ had similar values to the reference case, a much larger extent of the compositions was determined as capture origin rather than impact origin (Table \ref{tab:Dh_cases}).

% \begin{landscape}
\begin{figure*}
  \centering
  \includegraphics[bb=0 0 397 556]{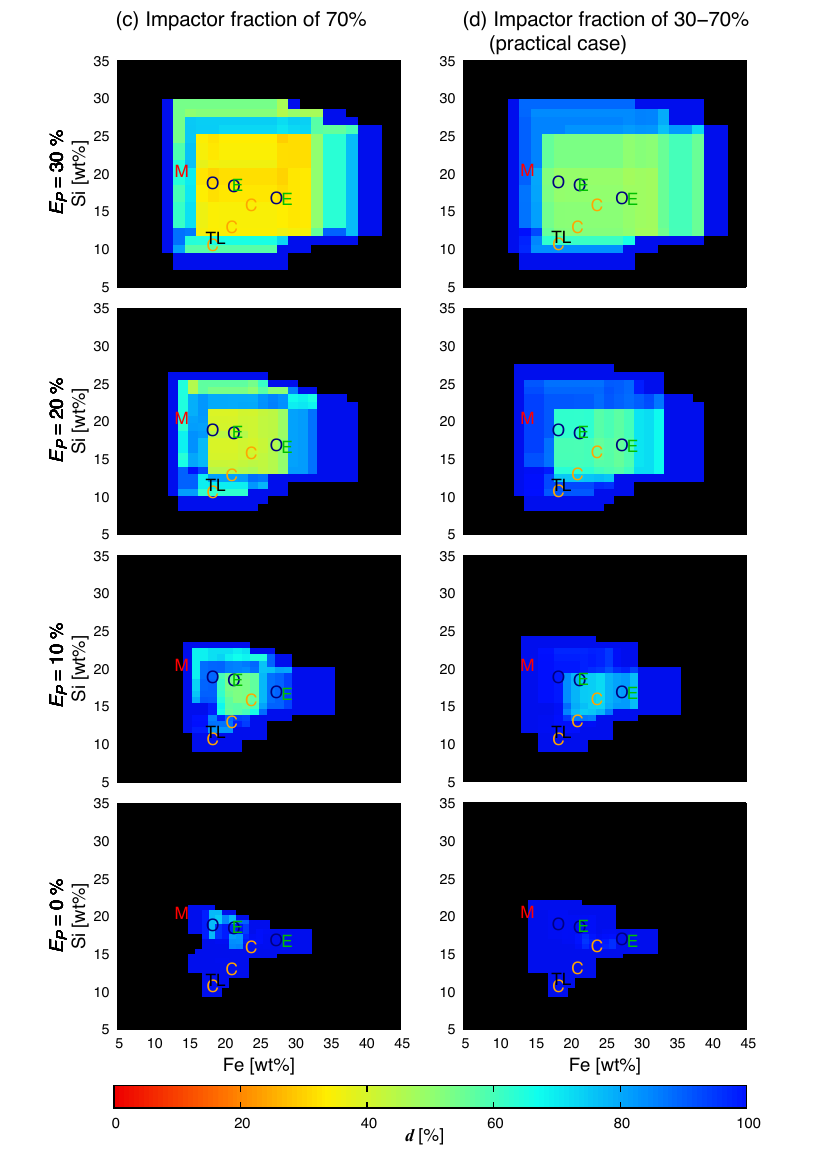}
  % \caption{Discrimination performance $d$ on Fe-Si diagrams when modeled asteroidal fraction for capture origin was 100\% and for impact origin was(a) 30\%, (b) 50\% (reference case), (c) 70\%, and (d) 30-70\% (practical case), along with end-members compositions (white label). Note that modeled asteroidal fractions were used as criteria when we judged the origin from the calculated mixing ratio (Section \ref{Criteria for capture/impact hypothesis}).}
\end{figure*}
% \end{landscape}
The distributions of $d$ in the practical case (Fig.~\ref{fig:d_dist}d) were similar to those in the reference case (Fig.~\ref{fig:d_dist}b). Compared to the reference case, the whole compositions were extended, resulting from the broad compositions assumed for the impact origin.

\section{Discussion}
\label{Discussion}
\subsection{Bulk Composition of Phobos and Discrimination Performance}
\label{Bulk composition of Phobos and Discrimination performance}
The inverse-solving calculations using our model revealed the relationship between the Phobos compositions measured by MEGANE and the reasonable origins. Here we discuss the relationship between the two-element compositions and the discrimination performances.

Two-element compositions had specific $d$ values. The majority of compositions with $d$ of 100\% showed either $d_{\mathrm{cap}}=$100\% or $d_{\mathrm{imp}}=$100\% (yellow and blue, respectively, in Fig.~\ref{fig:Two-elm_colors_multi}). In contrast, some compositions had $d=$ 100\%, $d_{\mathrm{cap}}<$100\%, and $d_{\mathrm{imp}}<$100\% (red in Fig.~\ref{fig:Two-elm_colors_multi}) and the proportion of such two-element compositions differ among the pairs of elements (Fig.~\ref{fig:Two-elm_colors_multi}), with the smallest for the pair of Fe-Si. This suggests that Fe-Si compositions best discriminate the origin of Phobos when only these 6 lithophile elements are considered. Additionally, Fe-Si compositions measured by MEGANE are also expected to have the smallest errors (\cite{lawrence_measuring_2019}).

Figure \ref{fig:d_dist}(a)($E_P = 30\%$) shows the overall trend that $d$ is larger when $\bm{P}$ is close to end-member compositions and smaller when $\bm{P}$ is an intermediate composition. However, it also shows that even if the composition completely agrees with an asteroidal composition, the origins are not always determined. For example, EL-like compositions ($\sim20\%$ Fe and $\sim19\%$ Si) can be explained by the mixture of Martian and some chondritic compositions.

\subsection{MEGANE Error and Discrimination Performance}
\label{MEGANE error and Discrimination performance}
The discrimination performance also depends on MEGANE's observation errors (see Section \ref{Reference case}). As $E_P$ decreased from 30\% to 0\%, $\mathscr{D}$ increased from approximately 60\% to more than 95\% in both the reference and practical cases (Table \ref{tab:Dh_cases}). MEGANE's instrumental performance and initial MMX operation plan estimates one-standard-deviation measurement uncertainties of 20\% for Fe, Si, and Th, and 33\% for O, Mg, and Ca (\citet{lawrence_measuring_2019}). In this case, $\mathscr{D}$ is approximated to 70\%, meaning that the origin will be determined from MEGANE observation in $\sim70\%$ of the possible cases considered in this study using only measurements of these 6 elements.

The observation errors for gamma-ray spectroscopy are strongly dependent on the total acquired measurement time and the altitude of the measurements. The relative precision of MEGANE's measurements can be improved if the MMX mission obtains MEGANE measurements beyond 10 days of total accumulated time at altitudes lower than the 1-body radius, under which are the conditions for which the current sensitivities were estimated (\citet{lawrence_measuring_2019}; \cite{nakamura_science_2021}). Recently, \citet{chabot_megane_2021} have investigated the potential footprints of MEGANE observations using the three-dimensional shape model of Phobos on the Small Body Mapping Tool (SBMT), as described in \citet{ernst_small_2018}. They have suggested that the MEGANE data resolution from the planned MMX trajectories approximates or is coarser than the independent spectral units on Phobos. Even with the larger MEGANE error ($E_P = 30\%$), more than 60\% of the compositional area derived a unique solution to the formation hypothesis.

The use of additional elements will also improve discrimination performances.
For example, \citet{lawrence_measuring_2019} suggested that MEGANE will measure the abundance of H, Na, K, Cl, and U in addition to the 6 elements used in this study. We specifically discuss the use of K abundance in Section \ref{The effect of volatile loss}. Since measurements of different element species have different measurement errors (\citet{lawrence_measuring_2019}), applying different errors for different elements in our model would be useful for more realistic estimates of the actual data analysis to determine the formation scenario, although the same MEGANE errors for all 6 elements were assumed in this study. The detailed MMX observation plan determined in the future will enable this model to be updated for a more specific discrimination performance.

\subsection{Asteroid-type Classification}
\label{Asteroid-type Classification}
One of the science goals of the MMX mission is to reveal the origin of the Martian moons to understand the processes for planetary formation and material transport (\citet{kuramoto_martian_2022}).
MEGANE's observations will help to achieve the MMX science goals and to distinguish between the capture and impact theories for Phobos's formation. MEGANE data have the added potential to reveal the type of asteroid related to the origin as well as the formation hypothesis. Previous studies also investigated the possibilities of asteroid or meteorite type identification using gamma-ray spectroscopy data. For example, \citet{prettyman_elemental_2012} showed with analysis of gamma-ray spectroscopic data acquired by the Dawn spacecraft that there is a consistency of the composition of 4 Vesta with HED meteorites, and \citet{peplowski_hydrogen_2015} also investigated the similarity between asteroid 433 Eros and L- or LL-chondrite compositions.

For the quantitative evaluation, we defined classification performance $\mathscr{C}$, $\mathscr{C}_{\mathrm{cap}}$, $\mathscr{C}_{\mathrm{imp}}$, which were given as % (Eqs.~\ref{eq:C}--\ref{eq:C_cap}). 
\begin{eqnarray}
  \label{eq:C}
  \mathscr{C}&=&\frac{n'_1+n'_2}{n_1+n_2}\times 100 \mathrm{\ [\%]}, \\
  \label{eq:C_cap}
  \mathscr{C}_{\mathrm{cap}}&=& \frac{n'_1}{n_1}\times 100 \mathrm{\ [\%]}, \\
  \label{eq:C_imp}
  \mathscr{C}_{\mathrm{imp}}&=& \frac{n'_2}{n_2}\times 100 \mathrm{\ [\%]}.
\end{eqnarray}

Note that $n'_1$ and $n'_2$ were the number of $\bm{P}$ classified into Case-1' and -2': $\bm{P}$ was explained (Case-1') by capture origin related to a unique asteroid type and (Case-2') by impact origin related to a unique impactor type. $\mathscr{C}$ represented the ratio of $\bm{P}$ which enabled the classification of asteroid type, while $\mathscr{C}_{\mathrm{cap}}$ and $\mathscr{C}_{\mathrm{imp}}$ only focused on capture and impact origins, respectively. Two-element classification performances ($c$, $c_{\mathrm{cap}}$, and $c_{\mathrm{imp}}$) were also calculated in the same way as described in Section \ref{Discrimination Performance}.

Furthermore, final performances ($\mathscr{F}$, $\mathscr{F}_{\mathrm{cap}}$, and $\mathscr{F}_{\mathrm{imp}}$) were defined as %(Eqs.~\ref{eq:F}--\ref{eq:F_imp}).

\begin{eqnarray}
  \label{eq:F}
  \mathscr{F} &=&\frac{\mathscr{D}}{100}\frac{\mathscr{C}}{100}\times 100=\frac{n'_1+n'_2}{n_1+n_2+n_3}\times 100 \mathrm{\ [\%]}, \\
  \label{eq:F_cap}
  \mathscr{F}_{\mathrm{cap}}&=&\frac{\mathscr{D}_{\mathrm{cap}}}{100}\frac{\mathscr{C}_{\mathrm{cap}}}{100}\times 100=\frac{n'_1}{n_1+n_2+n_3}\times 100 \mathrm{\ [\%]}, \\
  \label{eq:F_imp}
  \mathscr{F}_{\mathrm{imp}}&=&\frac{\mathscr{D}_{\mathrm{imp}}}{100} \frac{\mathscr{C}_{\mathrm{imp}}}{100}\times 100=\frac{n'_2}{n_1+n_2+n_3}\times 100 \mathrm{\ [\%]}.
\end{eqnarray}

\begin{figure}
  \centering
  \includegraphics[bb=0 0 386 412]{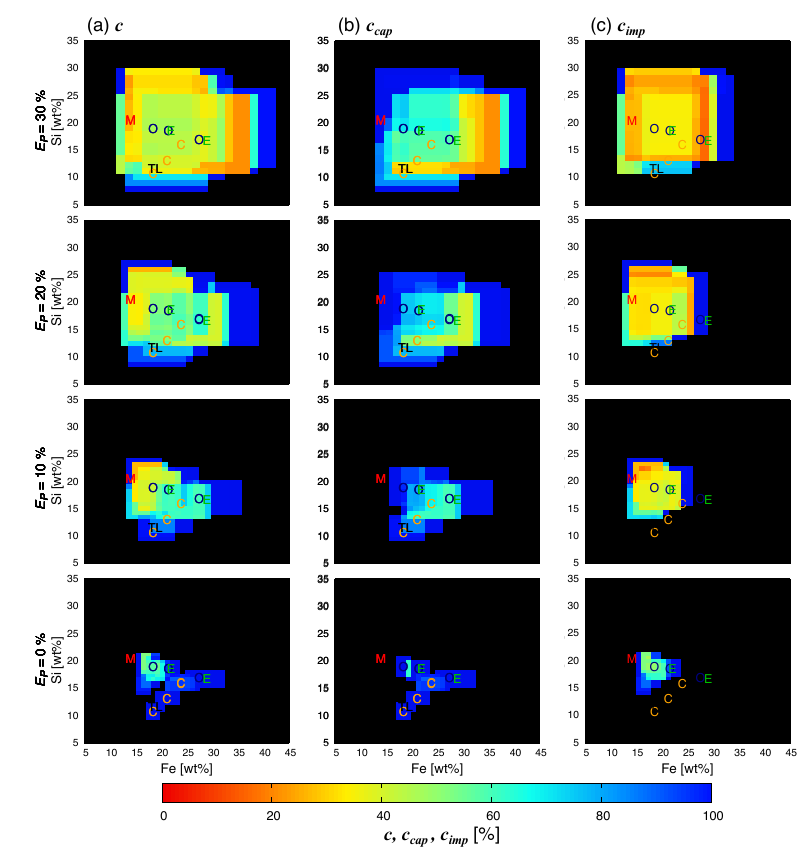}
  \caption{Classification performances (a) $c$, (b) $c_{\mathrm{cap}}$, and (c) $c_{\mathrm{imp}}$ on Fe-Si diagrams when modeled asteroidal fraction for impact origin was 50\% (reference case), along with end-member compositions (white label).}
  \label{fig:c_dist}
\end{figure}

For the reference case, $\mathscr{C}$ of approximately 40\% was derived (Table \ref{tab:C_cases}) when we assumed the present-expected MEGANE error ($E_P = 20\%$; \citet{lawrence_measuring_2019}), suggesting MEGANE's potential to classify the asteroid type with 40\% probability when the formation hypothesis is determined, based on these 6 lithophile elements alone. $\mathscr{C}$ improved to 42.9 and 74.9\% as $E_P$ decreased to 10 and 0\%. Comparison between $\mathscr{C}_{\mathrm{cap}}$ and $\mathscr{C}_{\mathrm{imp}}$ indicates the relative difficulty to classify the asteroid type in the case of the impact origin, which appears natural considering that the mixing with Martian materials decreases the compositional variations between different compositions of asteroids. The variation of $c$ showed a similar trend to that of $d$. $c$ was larger for compositions closer to end-member compositions than for the intermediate compositions (Fig.~\ref{fig:c_dist}a). $c_{\mathrm{cap}}$ and $c_{\mathrm{imp}}$ were larger for asteroid-like and Mars-like compositions (Figs.~\ref{fig:c_dist}b and c).

Final performance $\mathscr{F}$ was 37.4\% when $E_P = 20\%$, among which $\mathscr{F}_{\mathrm{cap}}$ and $\mathscr{F}_{\mathrm{imp}}$ were 23.5 and 13.9\%, respectively. When $E_P$ was changed between 0--30\%, $\mathscr{F}_{\mathrm{cap}}$ was always larger than $\mathscr{F}_{\mathrm{imp}}$ but they did not differ by a factor of 2.

In contrast to the reference case, the practical case calculation derived $\mathscr{F}_{\mathrm{imp}}$ larger than $\mathscr{F}_{\mathrm{cap}}$. This suggests that the asteroid classification is more difficult for the capture origin than for the impact origin because the compositional variation of $\bm{P}$ in the capture origin is smaller than that in the impact origin, where the uncertainty of the mixing ratio makes the compositional variation greater.

\begin{table}[h]
  \centering
  \begin{tabular}{cccccccccc}
    \toprule
    \multirow{3}{*}{$E_P \ [\%]$} & \multicolumn{4}{c}{Asteroidal fraction of 50\%} &                                   & \multicolumn{4}{c}{Asteroidal fraction of 30--70\%}                                                                                                                                           \\
                                  & \multicolumn{4}{c}{(reference case)}            &                                   & \multicolumn{4}{c}{(practical case)}                                                                                                                                                          \\
    \cmidrule{2-5} \cmidrule{7-10}
                                  & $\mathscr{C}$ [\%]                              & $\mathscr{C}_{\mathrm{cap}}$ [\%] & $\mathscr{C}_{\mathrm{imp}}$ [\%]                   & $\mathscr{F}$ [\%] &  & $\mathscr{C}$ [\%] & $\mathscr{C}_{\mathrm{cap}}$ [\%] & $\mathscr{C}_{\mathrm{imp}}$ [\%] & $\mathscr{F}$ [\%] \\
    \midrule
    30                            & 39.6                                            & 45.7                              & 32.8                                                & 25.6               &  & 34.7               & 64.7                              & 29.5                              & 22.1               \\
    20                            & 50.6                                            & 60.3                              & 39.8                                                & 37.4               &  & 47.5               & 72.2                              & 43.8                              & 35.9               \\
    10                            & 56.2                                            & 68.2                              & 42.9                                                & 48.6               &  & 51.0               & 76.0                              & 48.0                              & 44.7               \\
    0                             & 78.2                                            & 92.1                              & 65.8                                                & 74.9               &  & 66.5               & 92.9                              & 65.4                              & 65.2               \\
    \bottomrule
  \end{tabular}
  \caption{Classification performances ($\mathscr{C}$, $\mathscr{C}_{\mathrm{cap}}$, and $\mathscr{C}_{\mathrm{imp}}$; Eqs.~\ref{eq:C}--\ref{eq:C_imp}) calculated for modeled asteroidal fraction of 50\% (reference case) and 30--70\% (practical case).}
  \label{tab:C_cases}
\end{table}

Several sets of asteroid end-members cannot be separated in the 6-element compositional spaces because some chondrite groups have similar compositions. %Appendix? 
For example, the compositions of CC groups (especially CO, CK, CV, and CR) closely resemble each other, with a slight variation in Ca and Th abundances (Table \ref{tab:end-comp}). Such compositional similarity can make the determination of asteroid type difficult and reduce the value of $\mathscr{C}$. However, our model can more efficiently discriminate the groups within CC, OS, and EC classes.

\subsection{Limitations and Applications of the Mixing Model}
\label{Limitations and Applications of the Mixing Model}

\subsubsection{Formation scenarios not discriminated from MEGANE data}
\label{Formation scenarios not discriminated from MEGANE data}
Among the 24 (12 asteroid types $\times$ 2 hypotheses) formation scenarios assumed in our model, two combinations of formation scenarios were not adequately discriminated when only considering the abundance of 6 lithophile elements: the capture of an L-type asteroid vs the impact of an EL-type asteroid; and the capture of an H-type asteroid vs the impact of an EH-type asteroid. These scenarios result in the most similar elemental compositions of Phobos in our model, especially the two scenarios in the former combination result in extremely close compositions that agree within a relative difference of $<3\%$.

\subsubsection{Uncertainty of the Mixing Ratio}
\label{Uncertainty of the Mixing Ratio}
As shown in Section \ref{Practical case}, discrimination performances were dependent on the modeled asteroidal fraction for the impact origin. The actual mixing ratio will be estimated from laboratory analysis such as high-precision isotopic analysis of the returned samples from Phobos (\citet{usui_importance_2020}). However, MEGANE measurements will be performed several years before the sample analysis.
Therefore, the practical model examined the results with a more realistic range of 30--70\% for the mixing ratio (\citet{hyodo_impact_2017}) to compare the results with the reference case.
Since the mixing ratio will not yet be constrained when we obtain MEGANE observation data in the future, analysis with a wide range of possibilities will be needed. The practical model can be more useful in realistic MEGANE data analysis. Therefore, MEGANE data should be revisited after the sample analysis is completed using the sample-measured mixing ratio.

\subsubsection{The effect of volatile loss}
\label{The effect of volatile loss}
Volatile elements are considered key elements to discriminate the formation hypothesis because the giant impact may remove them from the impact-induced disk due to the higher temperature than the vaporization values (\citet{hyodo_impact_2018}). For example, thermodynamic calculations by \citet{pignatale_impact_2018} showed variations in the mineralogy of Phobos' building blocks depending on the disk temperature.
However, the effect of impact events on degassing from impact-induced disk materials is not yet fully understood and has not been examined in detail for consistency between the Moon-forming giant impact and the volatile depletion that has been investigated throughout lunar studies (\citet{canup_protolunar_2015}; \citet{charnoz_tidal_2021}; \citet{charnoz_evolution_2015}; \citet{karato_geophysical_2013}; \citet{krahenbuhl_volatile_1973}; \citet{nakajima_inefficient_2018}; \citet{pahlevan_equilibration_2008}; \citet{ringwood_komatiite_1987}; \citet{stevenson_origin_1987}). In this context, there should be a difference between observed volatile abundances and modeled volatile abundances for the impact origin scenario, since our mixing model does not take the preferential degassing into account. This would particularly affect K, given its volatile nature. Nevertheless, this does not mean that volatile elements are useless in our model. To demonstrate the performance of K abundances, discrimination performances were calculated using 7 elements (6 elements + K), assuming no loss of K during the impact formation process.

Adding K abundance resulted in $\mathscr{D}$ of 74.5, 83.7, and 96.8\% for $E_P$ of 30, 20, and 10\% when the modeled asteroidal fraction for the impact origin scenario was 50\% (reference case). Compared with the reference case (Table \ref{tab:Dh_cases}), $\mathscr{D}$ was improved by approximately 5--10\% for any $E_P$. $\mathscr{D}$ of 100\% was derived for $E_P$ of 0\%, which means that the origin can be completely distinguished if we precisely know the actual 7-element compositions of Phobos.
The improvement of $\mathscr{D}$ may be because the K abundance is sensitive to the contamination of Martian material due to the large variation in K abundance between Martian and asteroidal compositions.

Considering again the uncertainty of the effect of preferential volatile loss, the more reasonable and useful application of volatile data should be a forward-solving approach. After the formation scenario is determined from the inverse-solving using the 6 elements data, the modeled volatile abundance can be predicted by the forward-solving approach. By comparing the modeled abundance with the observed abundance, the degassing ratio will be derived. This will lead to a better understanding of the giant-impact event if that is the formation origin of the Martian moons.

\subsubsection{The effect of late accretion}
\label{The effect of late accretion}
Because Phobos has experienced a number of impact events after its formation, exogenic materials can deposit onto Phobos' surface as late accretion. % and present-day Phobos material may be composed not only of the two mixing end components but also of another impactor. 
Here we discuss the possibility to detect contamination of such exogenic materials.

A number of previous numerical studies (\citet{hyodo_transport_2019}; \citet{hyodo_searching_2021}; \citet{ramsley_mars_2013}; \citet{ramsley_origin_2013}; \citet{ramsley_stickney_2017}) have suggested that Martian materials ejected from Mars by asteroidal impacts should deposit on the surface of Phobos, regardless of its origin. %the impact conditions, because the orbital altitude of Phobos is small enough.
Thus, the returned samples acquired from the surface of Phobos by the MMX spacecraft are expected to contain these Martian materials, leading to the understanding of the habitability of Mars or providing a potential sign of life (\citet{fujita_assessment_2019}; \citet{hyodo_searching_2021}; \citet{kurosawa_assessment_2019}).

However, the concentration of Martian ejecta in Phobos regolith measurable by MEGANE is much smaller than the errors of the mixing ratio calculated with our model and much smaller than will be measurable by MEGANE.
\citet{ramsley_origin_2013} investigated the concentration of Martian ejecta delivered to Phobos by comparing the flux of Martian materials and that of the solar system projectile.
The present concentration on the Phobos regolith was estimated at $\sim250$ ppm within the surface $\sim 0.4$m layer. \citet{hyodo_transport_2019} updated the estimate by assuming the five largest impact craters on Mars as the source of Martian ejecta and derived a concentration of Martian ejecta and the solar system projectiles of approximately $\sim1000$ and $\sim10000$ ppm, respectively. The calculated errors were approximately 10--40\% in absolute values, which is orders of magnitude larger than the concentration of Martian materials (\citet{ramsley_origin_2013}; \citet{hyodo_transport_2019}). %In addition, the average contaminations of Martian and the solar-system materials can be further smaller within MEGANE's view, because MEGANE probes the elemental abundances to the depth of $\sim1$ m, deeper than the 0.4 m used in concentration estimates (\citet{ramsley_origin_2013}; \citet{hyodo_transport_2019}). 

For further application, our model can be applied to the mixing of more than 2 components. In this study, we assumed the mixing of only two end-member components: BSM and a chondritic component. However, our model is also applicable to other cases, such as the mixing of several asteroid compositions or the mixing between Mars and several impactors. The former is related to the case where the parent body of the captured asteroid was formed by the impact event of different types of asteroids and the latter to the case where several types of impactors formed the impact-induced disk from which Phobos was formed.

\section{Conclusion}
\label{Conclusion}
This study constructed a mixing model that connected the origin of Phobos and the elemental composition that will be measured by MEGANE for 6 lithophile elements (Fe, Si, O, Ca, Mg, and Th). Forward-solving predicts the elemental composition of the surface of Phobos from a given formation scenario. Inverse-solving discriminates the origin of the Martian moons from MEGANE observation data by calculating the mixing ratio of Martian and asteroid components.
The modeled performances to discriminate between formation hypotheses were calculated with varying the parameters of the mixing end-member compositions, the standard mixing ratio for the capture and impact origins, and the MEGANE observation error.

%The discrimination performances 
Our model shows that the ability to discriminate Phobos' origin scenario strongly depended on the MEGANE error. In a reference case, the origin was determined in 64.6\% of the whole compositional area, when the MEGANE error $E_P= 30\%$. As the observation error decreased to 20 and 10\%, the discrimination performances became larger to 73.8 and 86.6\%, respectively.
In the practical case, accounting for uncertainties during the impact event suggests a mixing ratio between 30 and 70\% for the impact origin. In this case, MEGANE data for these 6 lithophile elements can determine the origin with 70\% probability when $E_P =$ 20--30\%, which is suggested by the initial MMX plan.

As an additional application of our model, we found that MEGANE data may also be able to help classify the type of asteroid which was captured by or impacted Mars. The classification performance was approximately 50\% when $E_P =$ 20\%, which means that when the origin is determined from the compositional measurement, the asteroid type is also determined with a probability of 50\%, when these 6 lithophile elements are considered.

The use of other elements in the calculation improved the discrimination performance. For example, when we added another element K to the calculations, the performances were improved by 5--10\%. Note that due to the uncertainty of possible volatile loss from an impact-induced disk, the abundance of K and other volatile elements should be used for estimates of degassing ratios as well as providing insight into Phobos' origin.

This study identified the limitation of our mixing model for certain pairs of formation scenarios; for example, the capture of an L-type asteroid and the impact of an EL-type asteroid predict indistinguishably close compositions, making it difficult for MEGANE observation to discriminate between them. However, since these groups have distinct isotopic compositions, laboratory analysis of returned samples will discriminate the origin (\citet{fujiya_analytical_2021}; \citet{usui_importance_2020}).
As well as the sample collection, the MMX spacecraft will carry other scientific payloads such as a visible and near-infrared spectrometer, and a mass spectrometer (\citet{kuramoto_martian_2022}; \citet{usui_importance_2020}). Measurements by them will provide data complementary to MEGANE data.
The combination of all these observations will reduce the candidate formation scenarios, which will improve discrimination and classification performances. Observations during the MMX mission will comprehensively advance the understanding of the origin of Phobos.

\bibliography{Hirata_MEGANE_arXiv}

\end{document}